%% file: ms.tex
\documentclass[12pt,preprint]{aastex}

\slugcomment{Accepted for publication in ApJ}

\shorttitle{CO Mapping of IRAS~19312+1950}
\shortauthors{Nakashima et al.}

\begin{document}


\title{Wide Field CO Mapping in the Region of IRAS 19312+1950}


\author{Jun-ichi Nakashima\altaffilmark{1}, Dmitry A. Ladeyschikov\altaffilmark{2}, Andrej M. Sobolev\altaffilmark{2}, \\
Yong Zhang\altaffilmark{3,4}, Chih-Hao Hsia\altaffilmark{3,4,5}, Bosco H. K. Yung\altaffilmark{6}}

\altaffiltext{1}{Department of Astronomy and Geodesy, Ural Federal University,\\ Lenin Avenue 51, 620000, Ekaterinburg, Russia\\
Email(JN): nakashima.junichi@gmail.com}

\altaffiltext{2}{Astronomical Observatory, Ural Federal University,\\ Lenin Avenue 51, 620000, Ekaterinburg, Russia}

\altaffiltext{3}{Department of Physics, University of Hong Kong, Pokfulam Road, Hong Kong, China}

\altaffiltext{4}{Laboratory for Space Research, Faculty of Science, The University of Hong Kong, Pokfulam Road, Hong Kong, China}

\altaffiltext{5}{Space Science Institute, Macau University of Science and Technology, Avenida Wai Long, Taipa, Macau, China}

\altaffiltext{6}{N. Copernicus Astronomical Center, Rabia\'nska 8, 87-100 Toru\'n, Poland}


\begin{abstract}
We report the results of a wide field CO mapping in the region of IRAS 19312+1950. This IRAS object exhibits SiO/H$_2$O/OH maser emission, and is embedded in a chemically-rich molecular component, of which the origin is still unknown. In order to reveal the entire structure and gas mass of the surrounding molecular component for the first time, we have mapped a wide region around IRAS 19312+1950 in the $^{12}$CO $J=1$--0, $^{13}$CO $J=1$--0 and C$^{18}$O $J=1$--0 lines using the Nobeyama 45m telescope. In conjunction with the archival CO maps, we investigated a region with a size up to $20' \times 20'$ around this IRAS object. We calculated CO gas mass assuming the LTE condition, a stellar velocity against to the interstellar medium assuming an analytic model of a bow shock, and absolute luminosity using the latest archival data and trigonometric parallax distance. The derived gas-mass (225 M$_{\odot}$ -- 478 M$_{\odot}$) of the molecular component and the relatively large luminosity ($2.63\times10^{4}$ L$_{\sun}$) suggest that the central SiO/H$_2$O/OH maser source seems to be a red supergiant (RSG) rather than an asymptotic giant branch (AGB) star or post-AGB star.
\end{abstract}




\section{Introduction}
An isolated mid-infrared source, IRAS 19312+1950 (I19312, hereafter) exhibits SiO, H$_2$O and OH masers \citep{nak00a,nak07,nak11}, and the properties of the masers of this IRAS object are quite reminiscent of a mass-losing evolved star, such as asymptotic giant branch (AGB) stars, a post-AGB stars, and red supergiants (RSGs) \citep{nak11}. For example, exhibiting SiO maser emission is a typical characteristic of AGB stars and RSGs \citep{gen80,jew84,bar84,nak00b,ima02b,deg04,nak06,deg10,fok12}, a bipolar molecular jet traced in the H$_2$O maser line \citep{nak11} is quite similar to that found in oxygen-rich AGB/post-AGB stars \citep{ima02a,yun11}, and the intensity ratio of the OH maser satellite line to the main lines \citep[$>1$;][]{nak11} is consistent with that of mass-losing evolved stars with a cold dust envelope \citep{hab96,yun13,yun14}. Infrared properties of I19312 are also consistent with those of mass-losing evolved stars. For example, the mass-loss rate obtained by fitting dust radiative transfer models supports the AGB star status \citep{mur07}, and the near-infrared morphology is reminiscent of that of post-AGB stars and proto-planetary nebulae \citep[PPNe; ][]{deg04,mur07}.  In addition, \citet{Cooper13} classified I19312 as an M-type supergiant, since the near-infrared spectrum exhibits continuum emission that peaks in the middle of the 1.5 $\mu$m--2.5 $\mu$m region as well as showing molecular absorption lines owing to their cool atmospheres and have strong H$_2$O absorption. \citet{lum13} similarly classified I19312 as a PPN, based on the properties of infrared spectra and images.

At the same time, however, some other observational characteristics of I19312 cannot be explained in the standard scheme of the stellar evolution in its late stage. For example, the gas-mass estimated from a single-dish CO radio observation is 10~M$_{\odot}$ -- 15~M$_{\odot}$ \citep{nak04b,nak05}; such a large gas-mass is hardly explained as materials expelled from an AGB/post-AGB star (we give a new estimation of the gas mass in the present study; see, Section 4.1). Even if we assume an RSG with a larger initial-mass, the interpretation is not straightforward, because a rich-set of molecular species have been detected toward I19312 \citep{deg04}; it is not typical for RSGs. So far, more than a dozen of molecular species have been detected, including both carbon- and oxygen-bearing molecules \citep{deg04}; the complex chemistry including carbon-bearing molecules seems to be unusual for mass-losing evolved stars exhibiting SiO, H$_2$O and OH masers, because those molecules of masers are usually found only in oxygen-rich chemistry. The intensity peak velocities of the radio molecular lines of carbon-bearing molecules almost exactly corresponds to the systemic velocity obtained from maser observations \citep{deg04,nak11}. Interestingly, the detected molecules include CH$_3$OH \citep{deg04,nak15}, which has not been detected so far in any other evolved stars \citep[see, e.g.,][]{kaw95,cha97,cer00,he08,gom14}. Furthermore, Class I methanol lines have been detected toward I19312 recently \citep{nak15}; the authors suggested that Class I methanol maser emission is emitted from an interaction region between the outflow of an RSG and ambient molecular components.

Although a possibility of a young stellar object (YSO), which somehow exhibits SiO, H$_2$O and OH masers, has been considered in the past \citep[see, e.g.,][]{nak11}, the characteristics of I19312 are clearly different from such YSOs embedded in molecular clouds. For instance,  I19312 shows an isolated point-like feature in mid-infrared images \citep[see, e.g.,][]{nak11}, while all YSOs with SiO masers are deeply embedded in its natal clouds and therefore the background of such YSOs is generally quite bright at mid-infrared wavelengths \citep[e.g., Ori IRc 2, W51 IRs 2, and Sgr B2 MD5;][]{has86}; this is not the case for I19312. We would also note that we do not see any enhancement of the number density of stars in optical/near-infrared images (i.e., DSS, 2MASS, and UKDISS) around I19312. For these reasons, I19312 most likely cannot be explained as a YSO or a star at pre-main-sequence stages\footnote{We also note that a recent result of a near-infrared high-dispersion spectroscopy suggests that the $^{13}$C abundance is significantly enhanced compared to the $^{12}$C abundance toward the direction of the central star, and the $^{13}$C/$^{12}$C abundance ratio is far beyond the range of star forming regions (Parfenov, S. in private communications). }. 

We note that some groups working on massive YSOs have paid their attention on I19312, although no groups have provided a strong evidence for classifying I19312 as a YSO (three studies mentioned below picked out I19312 basically due only to a large flux densities in submm/mm continuum emission). \citet{Wienen12} have surveyed NH$_3$ emission toward a flux-limited sample of submm clumps detected by the APEX Telescope Large Area Survey: the GALaxy (ATLASGAL), which is an unbiased continuum survey of the inner Galactic disk at 870 $\mu$m. I19312 was included in their target list, and the NH$_3$ (1, 1) and (2, 2) lines were detected as \citet{deg04} already reported the detections of the same lines. The obtained rotational temperature (14.22 K) was consistent with that obtained in \citet{deg04} within uncertainty (the value obtained by Deguchi et al. was 19 K). \citet{Csengeri14} listed I19312 as one of the compact submm continuum sources that were identified in the ASTLASGAL project. The measured source size was $33'' \times 28''$ with a position angle of 121$^{\circ}$: the position angle is consistent with that of the molecular jet \citep[108$^{\circ}$--130$^{\circ}$; ][]{nak05,nak11}. \citet{shir13} searched the HCO$^+$ 3--2 and N$_2$H$^+$ 3--2 lines toward bright 1.1 mm continuum sources, which were found by the Bolocam Galactic Plane Survey \citep[BGPS; ][]{agu11}. Both lines were detected toward I19312, and this was the first detections of these lines.

So far, two hypotheses have been mainly considered for explaining I19312 \citep[see, e.g.,][]{nak11}: (1) a red nova formed by the merger of two main sequence stars (or two stars going to main sequence) and (2) a mass-losing evolved star, such as a post-AGB star or RSG, embedded in a small, isolated molecular cloud. However, both of two hypotheses have not been fully examined yet. One of the important viewpoints for considering the origin of I19312 is to see how the molecular gas of I19312 distributes and moves, particularly, in the outermost region of the nebulosity. The real extent and total mass of the molecular gas could be an important clue to distinguish the above two possibilities, because in the case (1) the total gas-mass cannot go beyond the total mass of merging stars (if we assume that additional gas does not flow into the merging system), while in the case (2) the total gas-mass can be much larger than the case (1). Through mapping observations in several molecular rotational lines \citep{nak04a,nak05}, we revealed that the central part of the nebulosity (within 15$''$--20$''$ from the central star) involves a spherically expanding component and a molecular bipolar outflow, while the behavior of the outermost nebulosity is still unknown. Although in the past we made two mapping observations with the Berkeley-Illinois-Maryland (BIMA) array and the Nobeyama Radio Observatory (NRO) 45~m telescope \citep{nak04b,nak05} in the CO rotational lines, those observations were not sufficient enough to reveal the entire distribution of the molecular component of I19312, because the BIMA observation resolved out the outermost nebulosity (the largest spatial frequency covered by the BIMA observation was roughly 15$''$--20$''$), and also because the mapping region of the NRO single-dish observation was too small (60$''$$\times$60$''$).

Thus, in order to reveal the entire structure of the molecular component of I19312, we made a wide field CO mapping in a region around I19312 using the 25-BEam Array Receiver System (BEARS) mounted on the NRO 45m telescope. We also investigated a wide field $^{13}$CO $J=1$--0 line image of the Boston University-Five College Radio Astronomy Observatory (FCRAO) Galactic Ring Survey (GRS) \citep{jac06}. In conjunction with both the new observation and archival data, we studied a region with a size up to $20' \times 20'$.


\section{Observations and Data Reduction}
The CO mapping observations in the region of I19312 were made in the period from 2013 May 23 -- Jun 2 using the NRO 45~m telescope (project number: CG122002). In total, 33 hours of the observing time were assigned in 11 days (3 hours per day), and roughly 21 hours out of the assigned 33 hours were usable for the observation; the remaining time were lost due to either mechanical problems or bad weather conditions. We observed three CO isotope lines:  i.e., $^{12}$CO $J=1$--0, $^{13}$CO $J=1$--0 and C$^{18}$O $J=1$--0. The rest frequencies used in the analysis were taken from \citet[][the values are given in Table~1]{lov92}. We used BEARS, which is a double-side band (DSB) superconductor-insulator-superconductor (SIS) heterodyne receiver array developed for the NRO 45~m telescope \citep{sun00}.  BEARS consists of $5 \times 5$ arrays with a fixed grid spacing of 41.1$''$ \citep{yam00}. The spectroscopic data were recorded by digital autocorrelators with a bandwidth of 512~MHz, 1024 frequency channels, and a frequency resolution of 500~kHz \citep{sor00}. The observed CO lines were placed at the center of the 512~MHz spectral window. The system temperature was ranged from 310~K to 550~K, depending on the observing frequency, weather conditions and the elevation of the telescope. The pointing accuracy was checked at the beginning and in the middle of each 3-hours observing session by observing the SiO $v=1$ and 2, $J=1$--0 lines of an AGB star V1111 Oph with the High-Electron-Mobility Transistor (HEMT) receiver H40; the pointing accuracy was typically within 1$''$--2$''$. We adopted the chopper-wheel method, switching between a room-temperature load and the sky, for primary intensity calibrations. This corrects for atmospheric attenuation and antenna ohmic losses, and converts the intensity scale to the antenna temperature in DSB [$T_{\rm A}^{*}$ (DSB)]. The beam size, main beam efficiency and aperture efficiency of the telescope were 14.7$''$--16.0$''$, 44\%--49\% and 30\%--36\%, respectively at 112 GHz (Note: BEARS consists of 25 arrays, and therefore the measurements of the beam size, main beam efficiency and aperture efficiency slightly vary from array to array; we used an averaged value in our analysis). 

We mapped a $500'' \times 500''$ square region in the $^{12}$CO $J=1$--0, $^{13}$CO $J=1$--0 lines and in a $250'' \times 250''$ square region in the C$^{18}$O $J=1$--0 line, using the on-the-fly (OTF) mapping technique \citep{saw08}. The regions were centered at the 2MASS position of I19312 (19h 33m 24.249s, +19$^{\circ}$ 56$'$ 55.65$''$, J2000.0), and the sides of the squares were set along the directions of the right ascension (X-direction) and declination (Y-direction). We scanned the regions along the X- and Y-directions several times in turn, and the maps in the two scan-directions were merged using the Basket-Weave technique \citep{eme88}. Scan numbers in each direction, rms.~noises achieved, velocity coverages, velocity resolutions are summarized in Table~1. The scan speed applied was 30$''$~s$^{-1}$ throughout all observations.

In order to reduce the OTF data, we used the software package NOSTAR, which was developed for handling BEARS data by NRO. We followed a standard reduction procedure instructed by NRO\footnote{see, http://www.nro.nao.ac.jp/\textasciitilde nro45mrt/html/obs/otf/index-e.html}; the standard procedure includes splitting spectra of each spectrometer, scaling the intensity, subtracting a baseline by fitting a low-order polynomial, flagging bad channels, applying the Basket-Weave technique, and creating FITS cubes. We applied the Bessel-Gaussian function to convolve the maps, using a grid spacing of 10$''$. The image processing and analysis of the reduced FITS cubes were made using the software package Miriad \citep{sau95}. 

In addition to the NRO data, we also used archival data of the Boston University-FCRAO Galactic Ring Survey (GRS). GRS is a molecular line survey of the inner Galaxy using the SEQUOIA multi-pixel array receiver on the FCRAO 14 m telescope in the $^{13}$CO $J=1$--0 line \citep{jac06}. The sensitivity, spectral resolution, angular resolution and sampling spacing of GRS are $<0.4$~K, 0.2~km~s$^{-1}$, 46$''$ and 22$''$, respectively. The location of I19312 is included in the two data cubes in the GRS archive\footnote{see, http://grunt.bu.edu/grs-stitch/download-all.php} (the file names of the used data in the archive are grs-55-cube.fits and grs-56-cube.fits).  The image processing and analysis of the GRS data were made using the software package Miriad.

\section{Results}
In Figures~1 and 2, we present the velocity-channel maps of the BEARS observations in the $^{13}$CO $J=1$--0 and $^{12}$CO $J=1$--0 lines. Although we observed the C$^{18}$O $J=1$--0 line as well, the velocity-channel map of this line is omitted due to the low signal-to-noise ratio. In Figures~1 and 2, we can clearly see the isolated CO emission of I19312 at the mapping center (see, 34.5~km~$^{-1}$--38.5~km~$^{-1}$ in Figure~1 and 35.5~km~$^{-1}$--39.5~km~$^{-1}$ in Figure~2). In the present observation, we revealed the entire structure of the CO emission of the molecular component of I19312 for the first time. As shown in the BIMA observation \citep{nak05}, the CO emission of I19312 shows a velocity gradient in the north-west to south-east direction  (see, the lower panel of Figure~3). The velocity gradient is also confirmed in Figure~1 (see, 34.5~km~$^{-1}$ -- 38.5~km~$^{-1}$), while in Figure~2, the gradient is not very obvious due to the low signal-to-noise ratio of the $^{12}$CO $J=1$--0 line and due also to the effect of self absorption \citep[see,][]{nak05}. In the upper panel of Figure~3, we show the velocity-integrated intensity (zero-moment) BEARS map of the $^{13}$CO $J=1$--0 line in the vicinity of I19312; the contour map is superimposed on the UKIDSS $J$, $H$, $K$-band composite color image. Comparing between the upper and lower panels in Figure~3, it is evident that the angular-size of the emission region seen in the BEARS map is roughly twice larger than that seen in the BIMA map; this difference of the emission region suggests that the non-negligible amount of the flux was resolved out in the BIMA observation due to interferometry \citep{nak05}. The angular-size of the emission region seen in the BEARS map is roughly 90$''$ in the north-south direction and 120$''$ in the east-west direction; this corresponds to $5.1 \times 10^{18}$~cm and $6.8 \times 10^{18}$~cm, respectively, at the distance of 3.8~kpc \citep{ima11}.

In Figure~3, we see that the intensity peak of the $^{13}$CO $J=1$--0 line is shifted roughly 10$''$ from that of the near-infrared emission in the north-west direction. This is due to the velocity gradient, which was already confirmed in the BIMA observation \citep{nak05}. Since the intensity of the blue-shifted component is slightly larger than that of the red-shifted component, the intensity peak of the zero-moment emission is shifted to the north-west direction, in which the flux of the blue-shifted component is dominant.

The upper panel of Figure~4 shows the spatially integrated BEARS spectra of I19312 in the $^{13}$CO $J=1$--0, $^{12}$CO $J=1$--0 and C$^{18}$O $J=1$--0 lines. The area of the integration is a circle with a diameter of 25$''$, which is centered at the 2MASS position of I19312. The intensity peak velocity of the $^{13}$CO $J=1$--0 of I19312 (at $\sim35$~km~s$^{-1}$) is consistent with the intensity peak velocity of other molecular lines \citep[roughly 35~km~s$^{-1}$ -- 36~km~s$^{-1}$; see, e.g.,][]{deg04,nak04b,nak15}, and is also consistent with the median of intensity peak velocities of SiO, H$_2$O and OH maser emission \citep[][Note: Since SiO, H$_2$O and OH masers exhibit multiple intensity peaks with a strong time-variation in its profile, it is difficult to obtain the precise systemic velocity of the maser source, but the median velocity of maser emission (i.e., the averaged velocity of the highest and lowest velocity components) is consistent with that of thermal lines within several km~s$^{-1}$]{nak11}. The intensity peak of the C$^{18}$O $J=1$--0 line also seems to be consistent with the velocity of other molecular lines, though it is slightly shifted ($\sim 1$~km~s$^{-1}$) due presumably to the low signal-to-noise ratio. Although the $^{13}$CO $J=1$--0 line exhibit 3 intensity peaks in its spectrum ($\sim29.3$~km~s$^{-1}$, $\sim35$~km~s$^{-1}$ and $\sim44.5$~km~s$^{-1}$), the $\sim29.3$~km~s$^{-1}$ and $\sim44.5$~km~s$^{-1}$ components seem to be parts of extended cloudlet, which is not directly related to I19312. The $^{12}$CO $J=1$--0 line does not show the intensity peak at $\sim35$~km~s$^{-1}$, which is the systemic velocity of the central maser source of I19312. As discussed in our previous papers \citep[see, e.g.,][]{nak04b,nak05}, this seems to be due to the self absorption in the $^{12}$CO $J=1$--0 line. 

An interesting point seen in Figures~1 and 2 is that we see an arc-like (or bow-shock-like) structure in the velocities from 34.5~km$^{-1}$ to 36.5~km$^{-1}$ in Figure~1 and from 35.5~km$^{-1}$ to 39.5~km$^{-1}$ in Figure~2, and it is reminiscent of interacting regions, which are occasionally found around mass-losing evolved stars \citep[see, e.g.,][]{uet06,mar07,jor11,dec12}.  In Figures~1 and 2, we indicated the crude location of the inner boundary of the arc-like structure by an ellipse, which is fit by eye inspection. We would like to note that this arc-like feature is also seen in the AKARI far infrared image \citep{kaw07}; see, Figure~5. The dashed straight lines in Figures~1 and 2 represent the long axis of the fitted ellipse. Interestingly, the position angle of the long-axis of the fitted ellipse (115$^{\circ}$) is close to that of the jet axes of the H$_2$O maser jet \citep[108$^{\circ}$;][]{nak11} and CO molecular bipolar flow \citep[130$^{\circ}$;][]{nak05}, and it is also not largely different from the position angle of the symmetric axis of the near-infrared structure \citep[143$^{\circ}$;][]{mur07}. Of course, this arc-like structure could be just extrinsic circumstances due to the inhomogeneity of the ambient gas and dust. However, as a possibility, such an arc-like structure of hydro dynamical interactions, can exist around a mass-losing evolved star. The detection of Class I methanol maser line \citep{nak15} is also suggestive of the existence of an interaction/shock region around I19312. Later in Section 4.2, we give some more quantitative discussions about this possible arc-like structure.

Figure~6 shows the GRS maps of the $^{13}$CO $J=1$--0 line, which covers a region wider than that of Figures~1 and 2 (the map size of Figure~6 is $20' \times 20'$). We note that, even though here we present only a part of the velocity coverage of the GRS data cube, almost no CO emission are detected out of the velocity range of Figure~6 (the entire velocity coverage of the GRS data cube is from $-5.0$~km~s$^{-1}$ to $85.0$~km~s$^{-1}$). In Figures~5 and 6, we compare CO maps, the IRAS point source catalog (PSC) sources and AKARI far-infrared images \citep{kaw07}, so that we could reinspect the isolation of I19312, which is mentioned in Section 1 and our previous papers \citep[see, e.g.,][]{nak11,nak15}. Within 10$'$ from the location of I19312, 7 IRAS PSC sources are found; the information of the nearby 7 IRAS sources are summarized in Table 2. Although we checked the SIMBAD database for these 7 IRAS sources, no significant information is found (except for IRAS measurements). In Figure~5 (AKARI far infrared image), we indicate the locations of the nearby 7 IRAS sources with the white crosses. As seen in Figure~5, except for IRAS 19309+1947, all other 6 IRAS sources do not have a clear counterpart in the AKARI far-infrared image; we see only extended cloudlet at the corresponding locations. In addition, the color of the cloudlet is significantly red in the image; this means that 160~$\mu$m emission is distinctly dominant (Note: 160~$\mu$m image represents red-color in the RGB color synthesis in Figure~5; the central wavelengths of the other two images used for the color synthesis are 90~$\mu$m [Green] and 60~$\mu$m [Blue]). In the Figure~6 (GRS $^{13}$CO map), we see $^{13}$CO emission at the locations of IRAS 19309+1947, IRAS 19308+1955 and IRAS 19306+1952. The GRS $^{13}$CO spectra of these three sources are given in the lower panel in Figure~4. The CO emission of IRAS 19308+1955 and IRAS 19306+1952 is relatively weak and seems to be shapeless rather than point-like. According to these properties, the IRAS sources in Table~2 (except for IRAS 19309+1947) are, most likely, not stellar-type objects, such as YSOs and dense cores, which must exhibit much clearer enhancements in CO and infrared flux densities. Presumably, what we are recognizing here as IRAS sources seems to be just the inhomogeneity of interstellar gas and dust. 

On the other hand, in Figure~5 we clearly see an infrared point source at the location very close to the IRAS position of IRAS 19309+1947. We mentioned this object in our previous study \citep{nak04b}. The IRAS $[12]-[25]$ color ($1.56-2.5 \log F_{12}/F_{25} = 2.66$; here, $F_{12}$ and $F_{25}$ are IRAS 12~$\mu$m and 25~$\mu$m flux densities) is typical as a YSO,  but the IRAS $[25]-[60]$ color ($1.88 - 2.5 \log F_{25}/F_{60}=5.32$; here, $F_{25}$ and $F_{60}$ are IRAS 25~$\mu$m and 60~$\mu$m flux densities) is out of the YSO range due to the large 60~$\mu$m flux \citep{str07}. A search of the 6.7 GHz methanol maser emission was negative toward this object \citep{szy00}. The intensity peak velocity of this object, which is found in the lower panel in Figure~4, is shifted roughly 6 km~s$^{-1}$ away from that of I19312. This velocity shift seems to be somewhat large when we compare the value with a typical velocity dispersion of a star cluster \citep[2~km~s$^{-1}$ -- 3~km~s$^{-1}$;][]{nak06,deg10}. At this moment, the true nature of this object is unclear, and moreover it is not clear whether this object is physically associated with I19312 or not.

\section{Analysis}
For better understanding the situation, in this section we obtain the following three values: (1) CO gas mass assuming the LTE condition, (2) a stellar velocity against to the interstellar medium assuming an analytic model of a bow shock, and (3) absolute luminosity using the latest archival photometric data and trigonometric parallax distance. 

\subsection{CO Gas Mass Assuming the LTE Condition}
One of the original purposes of the present CO observation is to reveal the entire structure of the molecular component of I19312 and to derive the total mass of it. Here, we estimate the CO gas mass of the isolated molecular component of I19312. In order to obtain the gas mass, we firstly created a map of the CO column density and then derived the mass by summing up the column density within the emission region. The detailed procedure of the calculation is as follows. 

Although we observed the $^{13}$CO $J=1$--0, $^{12}$CO $J=1$--0 and C$^{18}$O $J=1$--0 lines, we finally used only the $^{13}$CO $J=1$--0 for the mass estimation. The C$^{18}$O data was excluded due simply to its low signal-to-noise ratio. The $^{12}$CO $J=1$--0 line data was tentatively processed in the following way for trying to derive the excitation temperature, but we finally noticed that strong self-absorptions heavily disturbed the calculation as given the details later in this section, and did not use the data.  (Anyway, in what follows, we describe the details of the analysis processes both about the $^{13}$CO $J=1$--0 and $^{12}$CO $J=1$--0.)

Since the beam size, velocity resolution, and grid sizes of the original data cubes are different in each line, we convolved the data cubes, so that both maps have the same beam and grid sizes. The pixel numbers, grid size, velocity resolution, and beam size of the convolved data cubes are 95$\times$101$\times$150, 5$''$ $\times$ 5$''$,  1~km~s$^{-1}$, 9.55$''$, respectively. Then, we calculated the rms noise for each cube using emission free channels (specifically, we excluded velocities from 19.8~km~s$^{-1}$ to 49.8~km~s$^{-1}$ to avoid emission). The estimated rms noises are 0.33~K for $^{12}$CO and 0.10~K for $^{13}$CO.

We used the task \textsc{gaufit} in the Miriad software package \citep{sau95} to fit the emission feature of the $^{13}$CO $J=1$--0 and $^{12}$CO $J=1$--0 lines in each pixel with a single Gaussian function. In this Gaussian fitting, we assumed following three points: (1) the minimum FWHM of the line is 1~km~s$^{-1}$, (2) the minimum line intensity is 2~$\sigma$, (3) the central velocity of the Gaussian function is in the range from 30.5~km~s$^{-1}$ to 38.5~km~s$^{-1}$, which corresponds to the velocity range of the CO emission of of I19312.

Since the line intensity, line width and central velocity of the line were calculated for each pixel, the output of the \textsc{gaufit} task was given as 2-dimensional maps of the obtained values. Additionally, we further convolved the maps of the line intensity, line width and the central velocity of the line with a 10$''$ Gaussian beam to avoid artificial pixel-to-pixel sudden changes of the values.

After the Gaussian fitting, we calculated the column density distribution using the results of the Gaussian fitting. In principle, the excitation temperature ($T_{\rm ex}$) was obtained from the main beam antenna temperature ($T_{\rm a}$) of the $^{12}$CO line, using the solution of the radiative transfer equation \citep[equation 14.34 in][]{roh04}. However, as stated above, in the present case it is impossible to derive the excitation temperature from the $^{12}$CO $J=1$--0 due to the heavy self-absorption, which is lying over the systemic velocity of I19312. Thus, we assumed excitation temperatures for calculating the CO column density, so that we can estimate the mass only from the $^{13}$CO line; specifically, we calculated the CO column density about 10 different temperatures, which are ranged from 10~K to 100~K with a 10~K step.   

The optical depth of the $^{13}$CO $J=1$--0 line is obtained with the following equation \citep[equation 15.31 in][]{roh04}

$$ \tau(^{13}\mathrm{CO}) =  - \ln \left[ 1- \frac{T_\mathrm{B}^{13}}{5.3} \left\{  \left[ \exp \left(\frac{5.3}{T_\mathrm{ex}} \right)-1 \right]^{-1} - 0.16 \right\}^{-1} \right ], $$

\noindent
where $T_{\mathrm{B}}^{13}$ is the brightness temperature of the $^{13}$CO $J=1$--0 line, $T_\mathrm{ex}$ is the excitation temperature.  The source size of the compact component (90$''$-120$''$, see Section 3 for details) is larger than beam size (15$''$), thus we can assume $T_\mathrm{B}=T_\mathrm{mb}$.

For linear, rigid rotor molecules such as CO or $^{13}$CO, with the populations of all levels characterized by a single excitation temperature $T_{\rm ex}$, the column density $N(^{13}\mathrm{CO})$ and its optical depth $\tau$ are related as follows   \citep[equation 15.37 in][]{roh04} 

$$ N(^{13}\mathrm{CO})=3.0\times10^{14} \frac{T_\mathrm{ex}}{1-\exp(-5.3/T_\mathrm{ex})}  \int \tau^{13}(v)\mathrm{d}v ~ (\mathrm{cm}^{-2}). $$

In case of Gaussian line profile $\int \tau^{13}(v)\mathrm{d}v=\tau^{13}_0\sigma_v\sqrt{2\pi}$, where $\sigma_v$ -- velocity dispersion of $^{13}$CO line, $\tau_0$ -- optical depth in the center of the line. Column density of H$_2$ can be derived from abundance of $^{12}$CO/$^{13}$CO and CO/H$_2$:

$$ N(\mathrm{H}_2)=N(^{13}\mathrm{CO}) \times \mathrm{\frac{^{12}CO}{^{13}CO}}  \times \left[\mathrm{\frac{CO}{H_2}} \right]^{-1} ~ (\mathrm{cm}^{-2}).  $$

According to the value given in \citet{sim01}, we adopted the value of $\mathrm{CO}/{\rm H_2}~=~8 \times10^{-5}$. As mentioned in Section~1, the trigonometric parallax distance to I19312 is about 3.8~kpc \citep{ima11}; with this distance, the galactocentric distance of I19312 can be calculated to be 6.9~kpc. At this galactocentric distance, according to \citet{lan90}, the isotope ratio, $^{12}$CO/$^{13}$CO is $\simeq50$. Thus, for the calculation of the $\mathrm{H}_2$ column density, we used the value $^{13}\mathrm{CO}/\mathrm{H}_2~=~1.6 \times10^{-6}$. The mass is obtained by integrating the H$_2$ column densities over the source: 

$$ M=\mu m_\mathrm{H_2}\int{N_\mathrm{H_2}\mathrm{d}A}, $$

\noindent
where $\mu$ is the ratio of the interstellar gas mass to the mass of hydrogen molecule, $\mu\approx 1.33$ \citep{Hildebrand83}, $m_\mathrm{H_2}$ is the mass of hydrogen molecule. The surface element $\mathrm{d}A$ is related to the solid angle element $\mathrm{d}\Omega$ by $\mathrm{d}A=D^2\mathrm{d}\Omega$. If we combine previous three equations and constants, we can rewrite the equation for the gas mass in each pixel of the map with coordinate step  $\Delta\alpha$ and $\Delta\delta$:

$$ \frac{M}{M_\mathrm{\odot}} \simeq 5.02\times10^{-25} D_\mathrm{kpc}^2 \Delta\alpha^{\arcsec}\Delta\delta^{\arcsec} N_\mathrm{H_2}, $$

where $N_\mathrm{H_2}$ is the H$_2$ column density in each pixel of the map, $\Delta\alpha^{\arcsec}$ and $\Delta\delta^{\arcsec}$ -- map step in arcseconds, $D_\mathrm{kpc}$ -- distance to the source in kiloparsecs. We used following values for calculation: $D_\mathrm{kpc}=3.8$~kpc, $\Delta\alpha^{\arcsec}$ = $\Delta\delta^{\arcsec}$ = 5$\arcsec$. Mass is derived by summing the values from the each pixel of emission. 

According to the CO feature found in the maps, we calculated the mass over a circle with a 40$''$ radius, which is centered at the 2MASS position of I19312. Consequently, at the distance of 3.8 kpc, the mass of the source (the central isolated component of the I19312) is estimated to be in the range from 225 to 478 M$_{\odot}$; the range of the estimated mass corresponds to the range of the excitation temperature from 10 to 40~K (see, footnote\footnote{Although we calculated the mass in the range from 10 to 100~K as given in Table~3, the averaged excitation temperature of the molecular component of I19312 does not seem to go beyond 40~K (presumably, much less than 40~K), because the size of the molecular component is relatively large (5.1--6.8$\times 10^{18}$~cm), and also because the large size seems to prevent the heating the gas by the emission from the central star. Thus, here we give the mass corresponding to the temperature range of 10 -- 40~K as a representative value}). The results of the calculation are summarized in Table~3. One may think that the derived mass in the present study is far different from that estimated in previous studies \citep[10--15~M$_{\odot}$; see, e.g.,][]{nak04a,nak05}. Here, we would like to note that \citet{nak04a} derived a gas mass using the data of a single-point observation using a single-dish telescope (i.e., they did not obtain a map), and therefore a non-negligible amount of emission of the central molecular component was missed from their observation; Note that their beam-size (HPBW) was about 15$''$, while the integration region used in the present calculation is a circle with a 40$''$ radius as we mentioned above. Therefore, the mass derived by \citet{nak04a} is only a part of the mass derived in the present work.


\subsection{Stellar Velocity Assuming an Analytic Model of a Bow Shock}

As mentioned in Section~3, we found a bow-shock-like structure in velocity channels close to the systemic velocity of I19312 (see, Figures~1 and 2). If we assume that this feature is a real bow-shock, which is formed by hydrodynamical interaction between the interstellar medium and the stellar wind of a moving star, we can estimate the relative velocity between the interstellar medium and the star under some assumptions about the stellar wind parameters.

Assuming that the physical thickness of the shock shell is negligible and the ram pressure of the interstellar medium is balanced by that of the stellar wind, \citet{wil96} derived an analytic solution for the bow shock, which predicts that the distance between the apex and the central star ($R_0$) is given by the following formula.

$$R_0=\sqrt{\dot Mv_{\rm w}/(4\pi \rho_{\rm ISM} v_*^2)}$$

\noindent
Here, ${\dot M}$ is the mass-loss rate, $v_{\rm w}$ is the velocity of stellar wind, $\rho_{\rm ISM}$ is the mass density of the ISM, and $v_*$ is the velocity of star respect to the interstellar medium. The shape of the bow shock can be expressed by the following formula. 

$$R(\theta)=R_0\sin^{-1}\theta \sqrt{3(1-\theta\cot \theta)}$$

\noindent
Here, $\theta$ is the polar angle from the symmetric axis as seen from the central star at the coordinate origin. Based on the present CO observations, we derive an angular separation between the apex of the bow and the star of 125$''$. If we assume that the distance to this object is 3.8~kpc \citep{ima11} and the symmetry axis of the bow shock lies in the sky plane, we obtain $R_0=7.1\times10^{18}$~cm. Assuming a low density atomic interstellar medium ($n_{\rm ISM}=0.1$~cm$^{-3}$) and values for ${\dot M}$ and $v_{\rm w}$ \citep[$10^{-4}$~M$_\sun$yr$^{-1}$ and 25~km~s$^{-1}$, respectively,][]{nak04b,nak05}, we derive a stellar velocity of 11~km~s$^{-1}$.

The linear size $R_0=7.1\times10^{18}$~cm corresponds to a distance of roughly 2.3 pc. This seems to be a relatively large distance, because for matter ejected from a star it takes about $9.0 \times 10^4$ yrs to reach 
such a distance (assuming the expansion velocity is 25 km~s$^{-1}$) and the total mass ejected during this time (assuming the constant mass-loss rate 10$^{-4 \sim -5}$ M$_{\odot}$~yr) would be 90 M$_{\odot}$. For comparison, the distance is much larger than the estimated distance to bow shock in AGB stars is only about 0.08 pc in case of R Hya \citep{uet06} to about 0.3 pc in Betelgeuse \citep{dec12}. We note that a remnant of an AGB wind, of which the size is about 1.3 pc, has been detected toward a planetary nebula NGC 7293 \citep{Zhang12}; according to this size, it is not very unnatural even if hydrodynamical interactions are occurred at 1--2 pc away from the central star. If the central star of I19312 is/was a RSG as we discusss in Section 5, the mass-loss rate and the size of remnant stellar wind could be more immense.


\subsection{Absolute Luminosity}
In the past, we gave a couple of estimations about the absolute luminosity of I19312 \citep{nak04b,nak11}. However, the recent results of the trigonometric parallax measurement with the VLBI technique \citep{ima11} enables us to estimate a much more reliable value of the absolute luminosity, and furthermore many new photometric data have been released for the public. Here, we 
thoroughly collected photometric measurements of I19312, which are available for the public, and recalculated the absolute luminosity of I19312 using the most reliable distance information.

In order to construct the spectral energy energy distribution (SED) of I19312, we used the photometric data of the following observations, surveys, and archives: the optical INT/WFC Photometric H$\alpha$ Survey of the Northern Galactic Plane (IPHAS) $i\arcmin$-band magnitude, Two Micron All Sky Survey (2MASS), Near-infrared photometry at $J$, $H$, $K$ and $K_s$ bands from a ground-based observation \citep{nak04b}, Mid-infrared images of the object were extracted from the {\it Spitzer} Galactic Legacy Infrared Mid-Plane Survey Extraordinaire (GLIMPSE), Wide-Field Infrared Survey Explorer (WISE), Infrared Astronomical Satellite (IRAS) Sky Survey, Midcourse Space Experiment (MSX), Multiband Imaging Photometer of Spitzer (MIPS),  and AKARI IRC Survey. Many of the photometric data were obtained from the Data Discovery Service of the NASA/IPAC Infrared Science Archive\footnote{http://irsa.ipac.caltech.edu/cgi-bin/Radar/nph-estimation}. The {\it Spitzer}/GLIMPSE fluxes were measured using the method described in \citet{Zhang12} and \citet{Hsia14}. We used photometric data in the wavelength range between 0.763~$\mu$m and 2.096~mm (in total, 38 data points including upper limits).  The photometric measurements collected are summarized in Table~4, and the SED diagram is given in Figure~6.

The optical-infrared photometric measurements are corrected based on two interstellar extinction coefficients toward I19312: i.e., $A({\rm V})$= 16.3 \citep{Frerking82} and $A({\rm V})$ = 30.98 \citep{sch11}. The extinction coefficients $A({\rm V})$ were converted to $A({\rm J})$ using the conversion factor given by \citet{ste09}. Then, $A({\rm \lambda})$ of corresponding wavelengths were derived based on Table~1 in \citet{mat90}; we obtained the $A({\rm \lambda})$ value by interpolating the values in \citet{mat90} when the conversion factor was not found at the corresponding wavelength (the correction does not change the values at wavelengths longer than roughly 4~$\mu$m). 

The emerging flux of I19312 with extinction corrections are fitted by a three-component model using the same expression described in \citet{Hsia10}. As seen in Figure~6, only the shortest wavelength component (less than roughly $\lambda =$5~$\mu$m) is affected by the interstellar extinction correction. Since we have three cases of extinction corrections [i.e., $A({\rm V})$ = 30.98, 16.3, and 0], three temperatures of the fitted black bodies corresponding to each $A({\rm V})$ value are obtained: i.e., those are 13,000 $\pm$ 4,800 K [$A({\rm V})$ = 30.98], 2,100 $\pm$ 700 K [$A({\rm V})$ = 16.3], and 1100 $\pm$ 400 K [$A({\rm V})$ = 0]. The first result (13,000 K) seems to be clearly too high as a temperature of a stellar maser source. This presumably means that the large extinction coefficient [$A({\rm V})$ = 30.98] is not reliable. The dust component is roughly fitted by two black bodies with temperatures of 191 K and 47 K. With an assumed distance of 3.8 kpc \citep{ima11}, we derived the total luminosity by integrating the region under the three black body models. The derived total luminosities are $2.06\times10^{4}$ L$_{\sun}$ [$A({\rm V})$ = 0], $2.63\times10^{4}$ L$_{\sun}$ [$A({\rm V})$ = 16.3], and $1.54\times10^{6}$ L$_{\sun}$ [$A({\rm V})$ = 30.98]. For the reason, mentioned above, presumably the absolute luminosity must be close to $2.63\times10^{4}$ L$_{\sun}$. In fact, this value is close to those given in other studies: i.e., $9.6\times10^{3}$ L$_{\sun}$ at 3.1 kpc \citep{Cooper13} and $1.37\times10^{4}$ L$_{\sun}$ at 3.7 kpc \citep{lum13}.


\section{Discussion}
As mentioned in Section 1, the primary purpose of the present research is to reveal the entire structure and mass of the molecular component of I19312 to constrain the possibilities of the origin of this complex system, which consists of a stellar maser source and chemically-rich molecular component. In the present observation, we revealed the entire structure of the molecular component of I19312, and we estimated the CO gas mass of the molecular component. In this section, we interpret the observational results given in Section 3 based on the analysis given in Section 4.

\subsection{AGB/post-AGB stars and RSGs}

As given in Section 4, the gas mass estimated in the present study is roughly ranged from 225 to 478 M$_{\odot}$ (corresponding to the range of the excitation temperature from 10 to 40~K). Since we assumed the LTE and optically thin conditions for the $^{13}$CO $J=1$--0 line, the estimated gas mass should be considered as the lower limit. 

Our original question in the present research is whether the mass of the molecular component could be explained as materials expelled from a star (or a binary system consisting of stars). Although the bipolar molecular jet and extended infrared nebulosity of I19312 are reminiscent of those found in post-AGB stars and PPNe, which are the descendants of AGB stars, the large gas mass far beyond 200 M$_{\odot}$ never can be interpreted as materials expelled from a single AGB star, because the main-sequence mass of an AGB star is ranged roughly from 0.8 M$_{\odot}$ to 8.0 M$_{\odot}$ \citep[see, e.g.,][]{hab96,bus99}. Even if assuming a binary system consisting of 2--3 AGB stars, it is still impossible to explain such a large mass. Thus, with the large estimated gas mass, first of all, we can safely conclude that the molecular component of I19312 is not originated in a single AGB star (or even a binary system consisting of 2-3 AGB stars). If the central SiO/H$_2$O/OH maser source is originated in an AGB/post-AGB star, the surrounding molecular component must be pre-existing rather than materials expelled from the central star. If this is to the case, the situation would be explained that the central AGB/post-AGB star coexists in a small isolated cloudlet with a mass of 225 M$_{\odot}$ -- 478 M$_{\odot}$ and the star and the cloudlet share almost the same radial velocity of about 35~km~s$^{-1}$. However, since the age of an AGB star is typically several Gyr \citep[see, e.g.,][]{ols96,gua97,mac11}, presumably it is impossible to consider that the natal molecular gas is still remained in the vicinity of the star over such a long time scale. Therefore, if the central stars is really an AGB/post-AGB star, we have to consider that the coexisting situation between the star and surrounding gas is a product of "pure chance". 

Another major possibility of explaining a mass-losing evolved star must be an RSG. The observational properties of RSGs are, in many aspects, quite similar to those of AGB stars. For example, infrared colors of RSGs are basically the same with those of AGB stars, because the dust temperature is almost the same \citep[see, e.g.,][]{fok12}. RSGs exhibit a circumstellar envelope, which is often similar to that found in AGB/post-AGB stars in nature \citep[see, e.g.,][]{shi03}. Maser properties of RSGs are also similar to those of AGB stars, although the velocity ranges of maser emission of RSGs are somewhat larger than those of AGB stars \citep{nak07,deg10,fok12}. But, an important difference between RSGs and AGB/post-AGB stars are the initial mass. An RSG has a larger initial mass than an AGB star: its mass-range is roughly from 10 M$_{\odot}$ to 25 M$_{\odot}$ \citep{lev09}. And, therefore, the age of RSGs are much shorter than AGB stars: say, $<20$ million years \citep[see, e.g.,][]{dav07,dav08}. Even if we assume the initial mass of 10 M$_{\odot}$ -- 25 M$_{\odot}$, however, all the mass of the molecular component (225 to 478 M$_{\odot}$) cannot be explained as materials expelled from an RSG or a binary system including 2-3 RSGs (putting aside whether such a binary system is physically possible) as well as the case of AGB/post-AGB stars. Therefore, the molecular component, anyway, must be pre-existing even if we assume an RSG. However, since the age of RSGs are much younger than AGB stars, the surrounding materials could be still a natal cloud (or a remnant of natal cloud) for the central RSG. In fact, in some open clusters containing RSGs still posses molecular component, which could be a natal cloud of RSGs \citep{deg10}. In such a case, however, presumably there should be a concentration of stars, which are formed in the same natal cloud, but there is no such a concentration of stars found so far around I19312. [Note: in a high-resolution HST/WFC image, we find a weak concentration of several stars in the vicinity of a red star (within roughly 3.5$''$), which seems to be the source of SiO/H$_2$O/OH masers. But, it is not clear whether these stars have a physical relation with the central maser source.] In terms of the luminosity, the value derived in Section 4.3 ($2.63\times10^{4}$ L$_{\sun}$) is consistent with the case of an RSG \citep{woo83,gro09,fok12}, although it is also not very inconsistent with the upper limit of post-AGB stars \citep{rey01,whi01,van03}. For those reasons, at this moment, an RSG, which is embedded in the (remnant of) natal cloud, seems to be more preferable interpretation for I19312 compared to the case of AGB/post-AGB stars lying in a small molecular component by pure chance, although not all the observational properties are consistently explained yet.

\subsection{Other Possibilities and Related Issues}
In terms of the large mass of the molecular component, one may think that a mass-loss of a very massive star with an initial mass of 100 M$_{\odot}$ -- 200 M$_{\odot}$ could form a surrounding molecular component with a mass  nearly equivalent to the stellar mass. In fact, a dozen of stars with a quite large initial-mass more than 100~M$_{\odot}$ are known in the sky: for example, R136a1, R136a2, R136a3, NGC 3603-B, R136c, NGC 3603-C \citep{cro10}, HD~269810 \citep{wal95}, VFTS 682 \citep{vin12}, WR42e \citep{gva13}. However, the situation of such extreme massive stars seems to be far different from the case of I19312. The most importantly, such very massive stars definitely exhibit high-energy phenomena, such as X-ray emission and extremely high outflow-velocity, due to its large self gravity \citep[see, e.g.,][]{sew79,dam93,tow06}, developing an ionized region around the star \citep[see, e.g.,][]{oka03,jam04}. These are not the case for I19312: an ionized region has not been detected so far, for example, in a B${\gamma}$ imaging and free-free emission surveys \citep{nak04b,nak11}.

In our previous papers, we discussed a possibility of a merger of two main-sequence stars (or two stars going to the main-sequence) for explaining the situation of I19312 \citep[see, e.g.,][]{nak11}. Specifically, we pointed out some similarities between I19312 and V838 Mon \citep{nak11}: i.e., V838 Mon exhibits both SiO masers and extended nebulosity, which is detected in the CO radio lines, as well as I19312 does \citep{deg05,kam07,kam08}. Therefore, the most promising model explaining the observational properties of V838 Mon seems to be the merger of two stars \citep[see, e.g.,][]{sok07}. However, the present result suggests that the surrounding gas of I19312 (roughly, 225 to 478 M$_{\odot}$) is much larger than that of V838 Mon \citep[a few tens of solar masses,][]{kam08}, and such a large mass cannot be explained by the gas expelled from the two merging stars. Even in such a case, we still could assume that the SiO maser emission of I19312 are caused by the merger of two stars as well as V838 Mon, but to explain the merger, presumably we should see a dense stellar concentration around the merging stars, otherwise two stars have no chance (or very small chance) to meet to merge. In fact, \citet{afs07} found that V838 Mon is a member of a B-stars association and the narrow CO emission can be attributed to an interstellar medium within the cluster. However, such a clear star association is not found in the vicinity of I19312 so far, as partly mentioned in Section 5.1.

As we discussed in Section 5.1, if the molecular component of I19312 is a part of the natal cloud of an RSG with masers, another concern would be which parts of the CO emission found in Figures~1, 2 and 6 is the same natal cloud, which formed the RSG. As we calculated in Section 4.2, if we assume the arc-like feature seen in Figures~1 and 2 is caused by a bow-shock, the relative velocity is calculated to be 11 km s$^{-1}$. This relative velocity seems to be too large as a velocity dispersion of a single natal cloud, which forms RSGs. In fact, the velocity dispersion of an open cluster including RSGs are, at most, roughly $\sim$5 km~s$^{-1}$ \citep[typically, 2 km~s$^{-1}$ -- 3 km~s$^{-1}$;][]{nak06,deg10}. Therefore, if we believe that the bow shock is a real existence, the molecular gas lying in the south-east of I19312 is probably not directly related to the formation of the RSG in I19312.


\section{Summary}
We have reported the results of a wide field CO mapping in the region of I19312. We revealed the entire structure of the molecular component of I19312 for the first time. We calculated CO gas mass assuming the LTE condition, a stellar velocity against to the interstellar medium assuming an analytic model of a bow shock, and absolute luminosity using the latest archival photometric data and trigonometric parallax distance. The derived large gas-mass and relatively large luminosity suggests that the central SiO/H$_2$O/OH maser source seems to be an RSG rather than an AGB/post-AGB star, and the surrounding molecular component could be a natal cloud of an RSG.


\acknowledgments
This work is supported by Act 211 Government of the Russian Federation, agreement No. 02.A03.21.0006. YZ thanks the Hong Kong General Research Fund (HKU7073/11P) for the financial support of this study. Nobeyama Radio Observatory is a branch of the National Astronomical Observatory of Japan, National Institutes of Natural Sciences. We thank Sergey Y. Parfenov for stimulating discussions about the infrared high-despersion spectroscopy of IRAS 19312+1950 and the infrared sources in the surveyed region. We also thank Shuji Deguchi and Wayne Chau for their extensive help in the NRO observation.


\clearpage

\input{tab1.tex}

\clearpage

\input{tab2.tex}

\clearpage

\input{tab3.tex}

\clearpage

\input{tab4.tex}

\clearpage


\begin{figure}
\epsscale{1.0}
\plotone{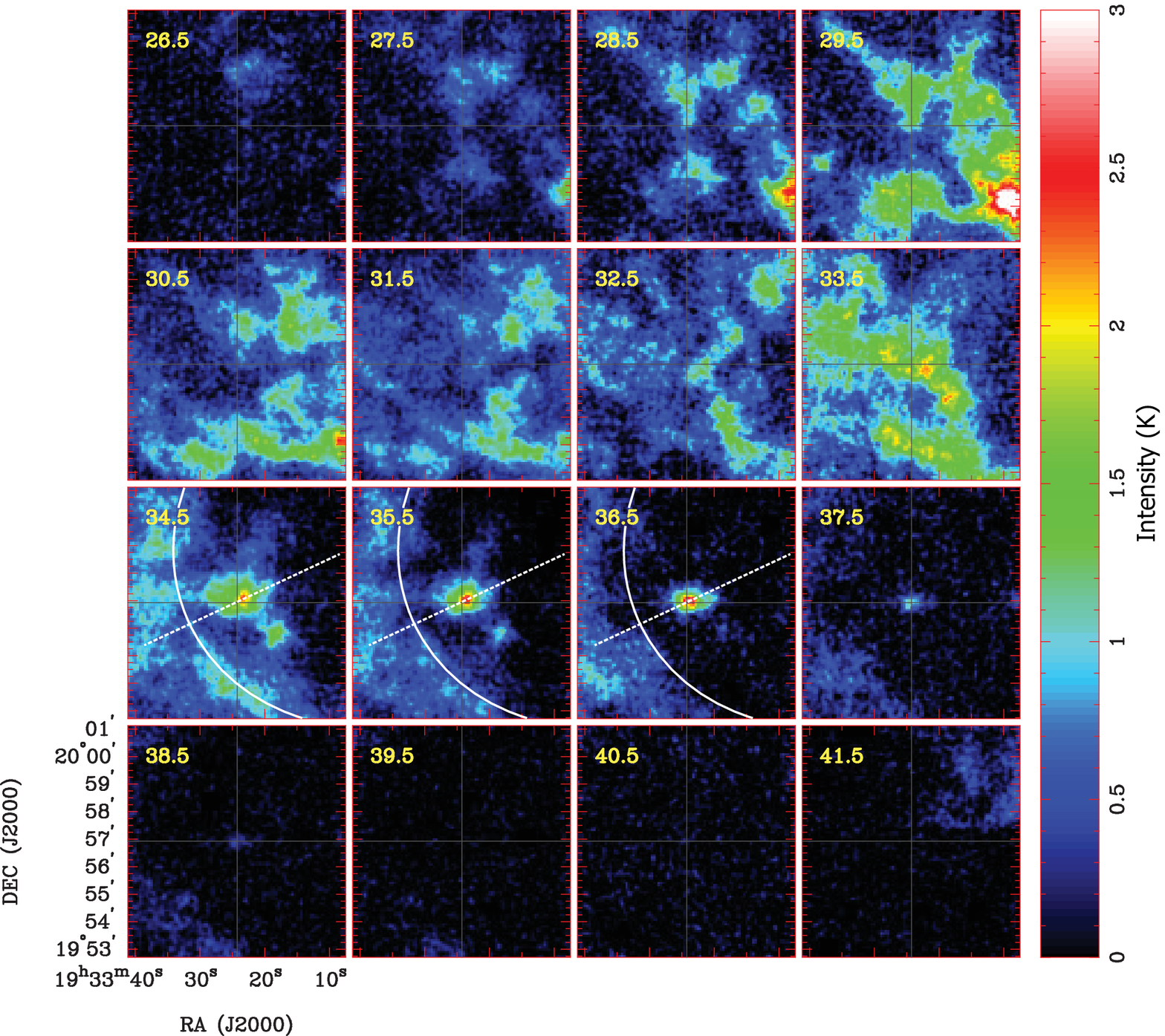}
\caption{Velocity-Channel maps of the $^{13}$CO $J=1$--0 line. The data were taken by the BEARS camera mounted on the NRO 45m telescope. The radial velocity ($V_{\rm lsr}$) is given in the upper-left corners of each panel. The color intensity scale in K is given on the right side of the channel maps. The intersection of the gray vertical and horizontal lines represents the location of IRAS 19312+1950 (2MASS position). The white curve (a part of an ellipse) indicates the inner boundary of a possible bow-shock feature, and the white broken line represents the long axis of the ellipse. \label{fig1}}
\end{figure}

\clearpage

\begin{figure}
\epsscale{1.0}
\plotone{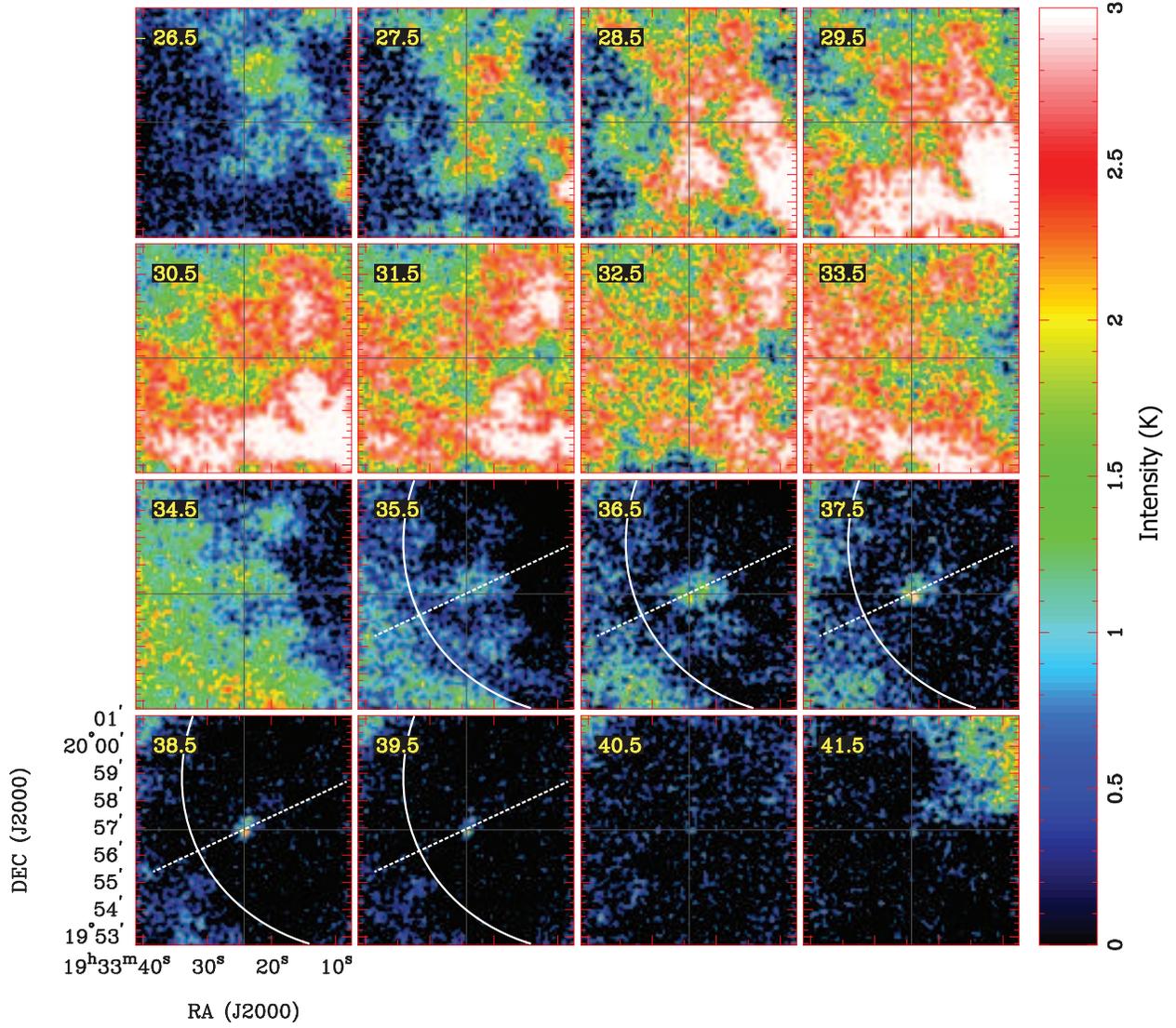}
\caption{Velocity-Channel maps of the $^{12}$CO $J=1$--0 line. The notations of the figure is the same as Figure~1. \label{fig2}}
\end{figure}

\clearpage

\begin{figure}
\epsscale{0.5}
\plotone{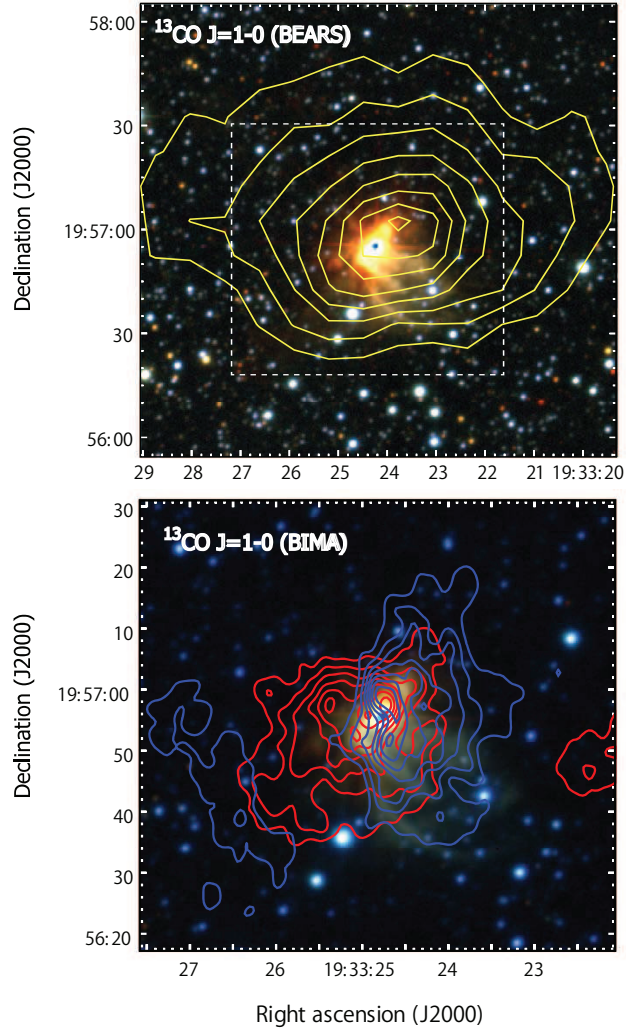}
\caption{{\bf Upper panel}: velocity-integrated intensity map of the $^{13}$CO $J=1$--0 line. The data was taken by the BEARS camera mounted on the NRO 45m telescope. The background is the $J$, $H$, $K$-bands composite color image of the UKIDSS data (both for the upper and lower panels). The velocity range for the integration is from 34.5~km~s$^{-1}$ to 39.5~km~s$^{-1}$. The contour levels start from 2.0~K~km~s$^{-1}$, and the levels are spaced every 0.871~K~km~s$^{-1}$. The highest contour corresponds to 8.1~K~km~s$^{-1}$. The white dashed square represents the size of the lower panel. {\bf Lower panel}: velocity-integrated intensity map of the $^{13}$CO $J=1$--0 line. The data were taken by the BIMA array \citep{nak05}. The blue contours represent the velocity ranges of 35--36~km~s$^{-1}$, and the red contours represent the velocity ranges of 37--38~km~s$^{-1}$. The contour levels start from 1.0~Jy~beam$^{-1}$, and the levels are spaced every 0.77~Jy~beam$^{-1}$.  \label{fig3}}
\end{figure}

\clearpage

\begin{figure}
\epsscale{0.5}
\plotone{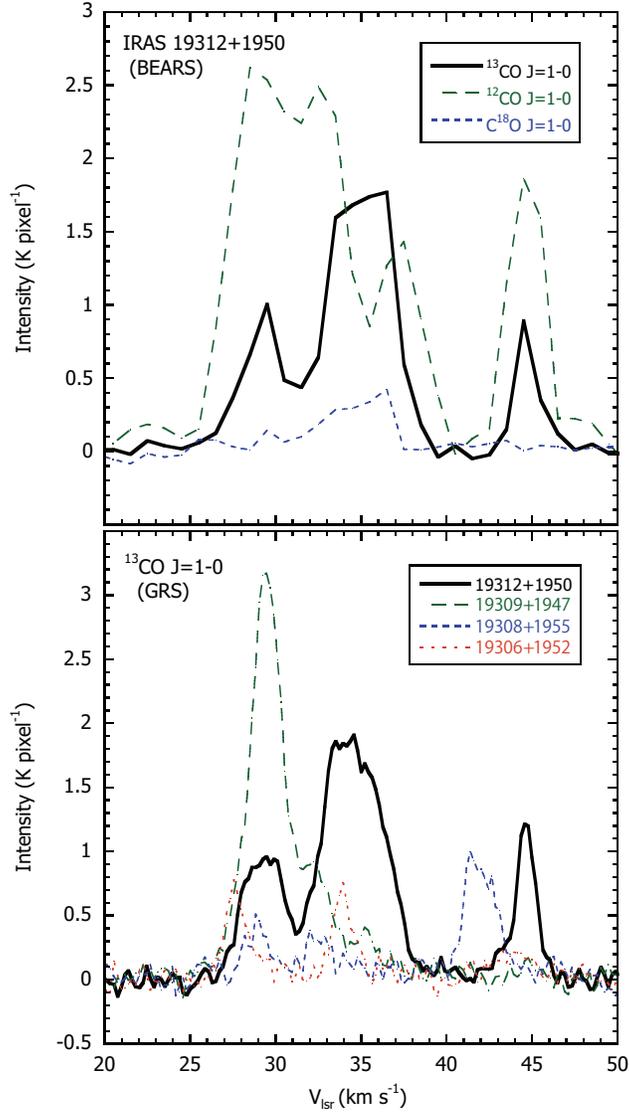}
\caption{{\bf Upper panel}:  Spatially integrated BEARS spectra of the $^{13}$CO $J=1$--0 (thick black line), $^{12}$CO $J=1$--0 (green broken) and C$^{18}$O $J=1$--0 (blue broken line) lines toward IRAS 19312+1950. The integration region is a circle with a radius of 25$''$, which is centered at the 2MASS position of IRAS 19312+1950. {\bf Lower panel}: Spatially integrated GRS spectra of the $^{13}$CO $J=1$--0 line toward IRAS 19312+1950 (thick black line),  IRAS 19309+1947 (green broken line), IRAS 19308+1955 (blue broken line) and IRAS 19306+1952 (orange dotted line).  The integration regions are circles with a radius of 25$''$, which are centered at each IRAS position (for IRAS 19312+1950 and IRAS 19309+1947, the 2MASS and AKARI positions are respectively used). \label{fig4}}
\end{figure}

\clearpage

\begin{figure}
\epsscale{0.6}
\plotone{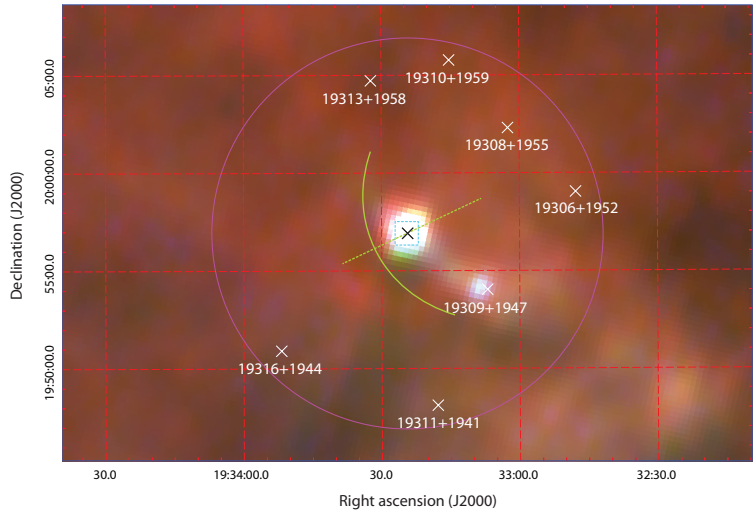}
\caption{Composite color image of the AKARI far-infrared images. The AKARI N60 (60~$\mu$m; blue), WideS (90~$\mu$m; green) and N160 (160~$\mu$m; red) images were combined to create the false RGB color. The black cross represents the location of I19312, and the white crosses represent the locations of nearby IRAS sources found within 10$'$ from I19312. The magenta circle with a radius of 10$'$, which is centered at I19312, is given as an indicator for the angular distance from I19312. The light green curve and broken line are the same with the white curve and broken line given in Figures~1 and 2 (i.e., the inner boundary of a possible bow-shock feature and the axis of the structure). The blue square corresponds to the white dashed square given in the upper panel in Figure 3 (i.e., the size of the BIMA map given in the lower panel in Figure~3). \label{fig5}}
\end{figure}

\clearpage

\begin{figure}
\epsscale{0.9}
\plotone{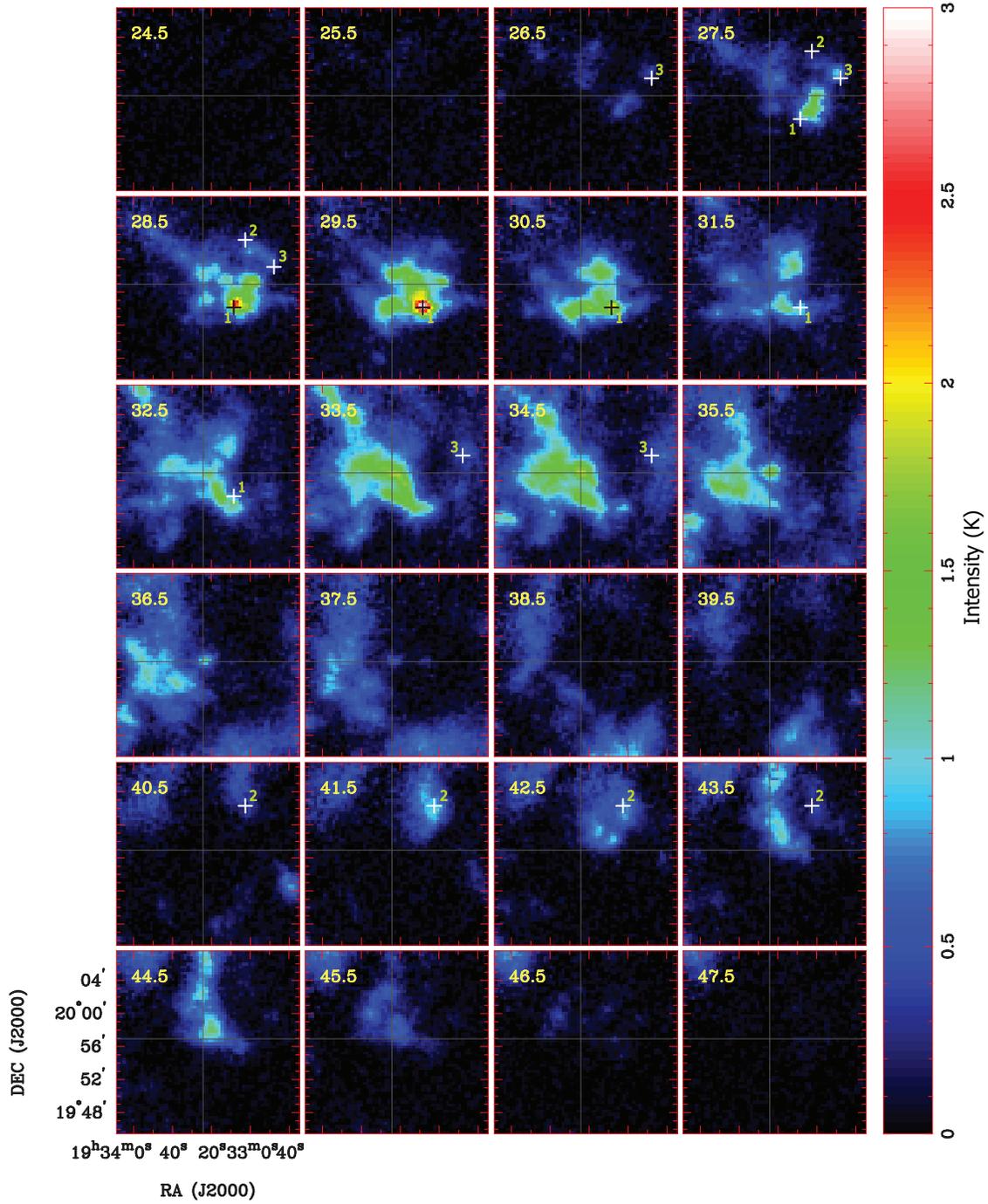}
\caption{Velocity-Channel maps of the $^{13}$CO $J=1$--0 line. The data were taken from the GRS data archive. The crosses with identification numbers indicate the location of IRAS point sources within 10$'$ from IRAS~19312+1950. The identification number 1, 2 and 3 corresponds to IRAS 19309+1947, IRAS 19308+1955 and IRAS 19306+1952. The other notations of the figure is the same as Figure~1. \label{fig6}}
\end{figure}

\clearpage

\begin{figure}
\epsscale{1.0}
\plotone{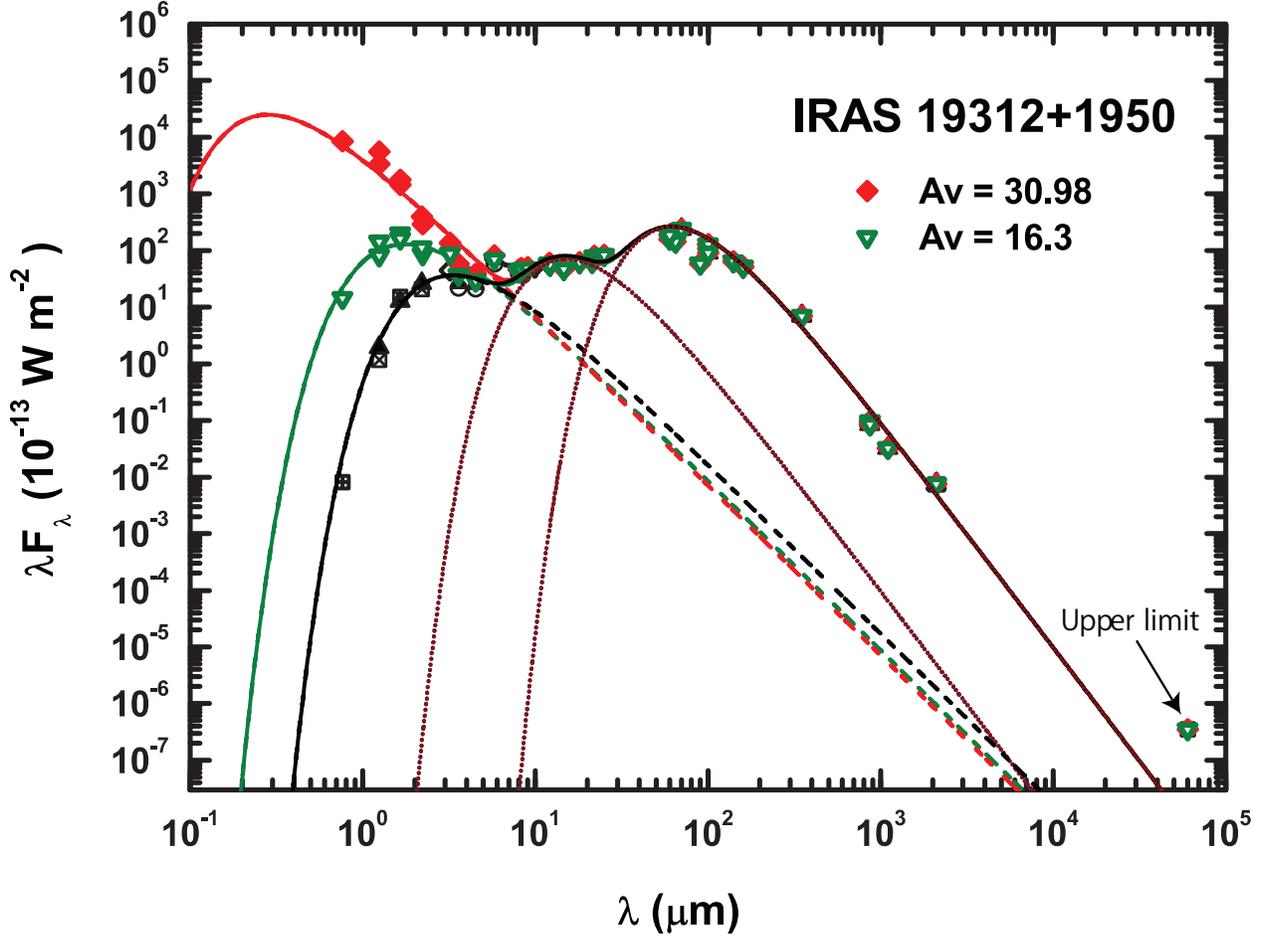}
\caption{Spectral energy distribution (SED) diagram of I19312 with extinction corrections in the wavelength range from 0.1 $\micron$ to 10 cm. The crossed square, crossed diamonds, filled triangles, open diamonds, half-filled circles correspond respectively to the data of IPHAS $i\arcmin$-band photometry, $J$, $H$, and $K$ near-infrared photometry, 2MASS, WISE, and Spitzer GLIMPSE survey (though, some of them may be unrecognizable due to overlapping with other symbols). The green open triangles and red filled diamonds respectively represent the photometric data corrected using the extinction values of $A({\rm V})$ = 16.3 and $A({\rm V})$ = 30.98. The black body curves are fitted to the emission from the photosphere (red, green, black lines) and two dust components (brown lines). The red, green and black lines represent the difference of the interstellar extinction corrections: i.e., $A({\rm V})$ = 30.98 (red), $A({\rm V})$ = 16.3 (green), and $A({\rm V})$ = 0 (black). \label{fig7}}
\end{figure}

\clearpage

\end{document}

%% file: tab1.tex
\begin{deluxetable}{lccc}
\tablecolumns{4}
\tablewidth{0pc}
\tablecaption{Summary of observational parameters}
\tablehead{
\colhead{} & \colhead{$^{12}$CO $J=1$--0} & \colhead{$^{13}$CO $J=1$--0} & \colhead{C$^{18}$O $J=1$--0} }
\startdata
Rest frequency$^{*}$ (GHz) & 115.271202 & 110.201354 & 109.782176 \\
Mapped region & $500''\times 500''$ & $500''\times 500''$ & $250'' \times 250''$ \\
Scan number (X-direction) & 3 & 12 & 5 \\
Scan number (Y-direction) & 2 & 11 & 4 \\
Velocity coverage (km~s$^{-1}$) & $\pm$666.3 & $\pm$696.9 & $\pm$700.0 \\
Velocity resolution (km~s$^{-1}$) & 1.30 & 1.36 & 1.37 \\
r.m.s. noise (K) & $2.13 \times 10^{-1}$ & $6.05 \times 10^{-2}$ & $9.85 \times 10^{-2}$ \\
\enddata
\tablenotetext{*}{Rest frequencies are taken from \citet{lov92}.}
\end{deluxetable}

%% file: tab2.tex
\begin{deluxetable}{lccccc}
\tablecolumns{6}
\tablewidth{0pc}
\tablecaption{IRAS PSC souces within 10$'$ from IRAS 19312+1950}
\tablehead{
\colhead{IRAS name} & \colhead{$F_{12}$} & \colhead{$F_{25}$} & \colhead{$F_{60}$} & \colhead{Separation} & \colhead{CO emission}\\
\colhead{} & \colhead{(Jy)} & \colhead{(Jy)} & \colhead{(Jy)} & \colhead{($''$)} & \colhead{km~s$^{-1}$}}
\startdata
19306+1952 & 1.40 &  0.49 & 3.93 & 562 & 26.5--28.5, 33.5--34.5\\
19308+1955 & 1.00 &  0.59 & 3.93 & 459 & 27.5--28.5, 40.5--43.5\\
19309+1947 & 1.22 & 15.40 & 366.10 & 291 & 27.5--32.5 \\
19310+1959 & 1.71 &  1.08 & 1.87 & 547 & no CO emission \\
19311+1941 & 0.62 &  1.23 & 4.44 & 537 & no CO emission \\
19313+1958 & 0.96 &  0.39 & 4.00 & 490 & no CO emission \\
19316+1944 & 0.56 &  0.42 & 2.79 & 548 & no CO emission \\
\enddata
\tablecomments{$F_{12}$, $F_{25}$ and $F_{60}$ are IRAS flux densities at 12~$\mu$m, 25~$\mu$m and 60~$\mu$m.}
\end{deluxetable}

%% file: tab3.tex
\begin{deluxetable}{ccccc}
\tablecolumns{2}
\tablewidth{0pc}
\tablecaption{Results of the LTE Mass Estimation}
\tablehead{
\colhead{T$_{\mathrm{ex}}$} & \colhead{$\tau(^{13}\mathrm{CO})$} & \colhead{$N(\mathrm{CO})$} & \colhead{$N(\mathrm{H_2})$} & \colhead{Mass}\\
\colhead{(K)} & \colhead{} & \colhead{$10^{16}$ cm$^{-2}$} & \colhead{$10^{22}$ cm$^{-2}$} & \colhead{(M$_{\odot}$)}}
\startdata
\phantom{0}10 &  0.32 & 1.02 &  0.63 & 225 \\
\phantom{0}20 &  0.12 & 1.31 & 0.82  & 290 \\
\phantom{0}30 &  0.07 & 1.73 & 1.08 & 382  \\
\phantom{0}40 &  0.05 & 2.16 & 1.35 & 478  \\
\phantom{0}50  &  0.04 & 2.61 & 1.62 & 576 \\
\phantom{0}60  &  0.03 & 3.05 & 1.91 & 674 \\
\phantom{0}70  &  0.028 & 3.50 & 2.19 & 773 \\
\phantom{0}80  &  0.024 & 3.95 & 2.47 & 872 \\
\phantom{0}90  & 0.018 & 4.40  & 2.75 & 971 \\
\phantom{0}100 & 0.016 & 4.84 & 3.03 & 1070 \\
\enddata
\tablecomments{T$_{\mathrm{ex}}$: excitation temperatures assumed, $\tau(^{13}\mathrm{CO})$: mean $^{13}$CO optical depths of the source, $N(\mathrm{CO})$ and $N(\mathrm{H_2})$: column densities of CO and H$_2$, respectively, Mass: derived masses.}
\end{deluxetable}

%% file: tab4.tex
\begin{deluxetable}{lll}
\tabletypesize{\scriptsize}
\tablecaption{Photometric Measurements of IRAS 19312+1950}
\tablewidth{0pt}
\tablehead{
\colhead{Measurements (flux unit)} & \colhead{Flux values} & \colhead{Reference} \\
\hline
\multicolumn{3}{c}{Optical data}
}
\startdata
IPHAS i$\arcmin$ (mag) & 18.11$\pm$0.04 & \citet{Barentsen14} \\
\hline
 \multicolumn{3}{c}{Infrared data} \\
\hline
2MASS J (mag) & 11.332$\pm$0.029 & \citet{Cutri03} \\
2MASS H (mag) & 7.718$\pm$0.024 & \citet{Cutri03} \\
2MASS K$_{s}$$^a$ (mag) & 6.615 & \citet{Cutri03} \\
J (mag) & 10.7$\pm$0.1 & \citet{nak04b} \\
H (mag) & 7.80$\pm$0.1 & \citet{nak04b} \\
K (mag) & 6.30$\pm$0.1 & \citet{nak04b} \\
WISE 3.4 $\micron$ (Jy) & 4.365$\pm$0.076 & \citet{Cutri12} \\
WISE 12 $\micron$ (Jy) & 0.253$\pm$0.01 &  \citet{Cutri12} \\
WISE 22 $\micron$ (Jy) & 2.122$\pm$0.002 &  \citet{Cutri12} \\
Spitzer 3.6 $\micron$$^b$ (Jy) & 2.65$\pm$0.46 & This study \\
Spitzer 4.5 $\micron$$^b$ (Jy) & 3.21$\pm$0.58 & This study \\
Spitzer 5.8 $\micron$$^b$ (Jy) & 11.42$\pm$1.67 & This study \\
Spitzer 8.0 $\micron$$^b$ (Jy) & 10.52$\pm$1.33 & This study \\
MSX 8.3 $\micron$ (Jy) & 11.84$\pm$0.49 & \citet{Egan03} \\
MSX 12.1 $\micron$ (Jy) & 22.68$\pm$1.13 & \citet{Egan03} \\
MSX 14.7 $\micron$ (Jy) & 23.08$\pm$1.41 & \citet{Egan03} \\
MSX 21.3 $\micron$ (Jy) & 45.00$\pm$2.70 & \citet{Egan03} \\
IRAS 12 $\micron$ (Jy) & 22.61$\pm$0.95 & \citet{Moshir92} \\
IRAS 25 $\micron$ (Jy) & 69.77$\pm$2.73 & \citet{Moshir92} \\
IRAS 60 $\micron$ (Jy) & 317.60$\pm$17.15 & \citet{Moshir92} \\
IRAS 100 $\micron$ (Jy) & 427.00$\pm$21.35 & \citet{Moshir92} \\
AKARI 9 $\micron$ (Jy) & 13.71$\pm$0.14 & \citet{Ishihara10} \\
AKARI 18 $\micron$ (Jy) & 37.02$\pm$0.45 & \citet{Ishihara10} \\
AKARI 65 $\micron$ (Jy) & 313.10$\pm$29.40 & \citet{Ishihara10} \\
AKARI 90 $\micron$ (Jy) & 176.30$\pm$38.90 & \citet{Ishihara10} \\
AKARI 140 $\micron$ (Jy) & 302.90$\pm$42.00 & \citet{Ishihara10} \\
AKARI 160 $\micron$ (Jy) & 275.10$\pm$31.80 & \citet{Ishihara10} \\
IRAS IGA 60 $\micron$ (Jy) & 379.62$\pm$0.68 & \citet{Mottram10} \\
Spitzer MIPS 70 $\micron$ (Jy) & 556.13$\pm$3.07 & \citet{Mottram10} \\
IRAS IGA 100 $\micron$ (Jy) & 303.13$\pm$6.55 & \citet{Mottram10} \\
ATLASGAL 870 $\micron$ (Jy) & 2.79 & \citet{Csengeri14} \\
ATLASGAL F 870 $\micron$ (Jy) & 2.45 & \citet{Wienen12} \\
BGPS 1100 $\micron$ (Jy) & 1.18$\pm$0.16 & \citet{Dunham11} \\
BGPS 1100 $\micron$ (Jy) & 1.186$\pm$0.21 & \citet{Rosolowsky10} \\
\hline
 \multicolumn{3}{c}{Radio data} \\
\hline
Planck 857 GHz (mJy) & 98525$\pm$8330.75 & \citet{Planck14} \\
Planck 143 GHz (mJy) & 802.97$\pm$132.53 & \citet{Planck14} \\
RMS 5 GHz$^c$ (mJy) & 0.7 & \citet{Cooper13} \\
\enddata
\tablenotetext{{\it a}}{Note that the 2mass K$_{s}$ band flux is unreliable.}
\tablenotetext{{\it b}}{Measured from Spitzer GLIMPSE images.}
\tablenotetext{{\it c}}{Note that the flux measured from 5 GHz is an upper limit.}
\label{tab1}

\end{deluxetable}